\documentclass[]{aastex631}

\usepackage{mathtools}
\usepackage{soul}
\usepackage{fontawesome}
\usepackage{esvect}
\usepackage{color}
\usepackage[normalem]{ulem}

\newcommand\given[2]{\left(#1\;\middle|\;#2\right)}

\newcommand{\msun}{\mathrm{M}_{\odot}}

\newcommand{\boldSigma}{\boldsymbol{\Sigma}}
\newcommand{\boldSigmaErr}{\boldsymbol{\Sigma}_{\rm err}}

\newcommand{\boldX}{\boldsymbol{x}}
\newcommand{\boldY}{\boldsymbol{y}}
\newcommand\hatboldY{\widehat{\boldsymbol{y}}}
\newcommand\hatY{\widehat{y}}

\newcommand\boldb{\boldsymbol{\beta}}

\newcommand\expymu{\langle\boldsymbol{y} \rangle_{\mu}}

\newcommand{\MStar}{M_{\rm \star,\,tot}}
\newcommand{\MGasHot}{M_{\rm gas,\,hot}}
\newcommand{\ctwohc}{c_{\rm 200c}}
\newcommand{\Mtwohc}{M_{\rm 200c}}
\newcommand{\Rtwohc}{R_{\rm 200c}}

\definecolor{bleudefrance}{rgb}{0.19, 0.55, 0.91}

\definecolor{purple}{RGB}{128, 0, 128}

\shorttitle{KLLR method}
\shortauthors{A. Farahi et al.}

\begin{document}

\title{\textsc{KLLR}: A scale-dependent, multivariate model class for regression analysis}

\correspondingauthor{Arya Farahi}\email{arya.farahi@austin.utexas.edu}

\author[0000-0003-0777-4618]{Arya~Farahi}
\affil{Departments of Statistics and Data Science, University of Texas at Austin, Austin, TX 78757, USA}
\author[0000-0003-3312-909X]{Dhayaa~Anbajagane}
\affil{Department of Astronomy and Astrophysics, University of Chicago, Chicago, IL 60637, USA}
\affil{Kavli Institute for Cosmological Physics, University of Chicago, Chicago, IL 60637, USA}

\author[0000-0002-4876-956X]{August~E.~Evrard }
\affil{Departments of Physics and Astronomy, Leinweber Center for Theoretical Physics, University of Michigan, Ann Arbor, MI 48109, USA}

\begin{abstract}
The underlying physics of astronomical systems governs the relation between their measurable properties. Consequently, quantifying the statistical relationships between system-level observable properties of a population offers insights into the astrophysical drivers of that class of systems. 
While purely linear models capture behavior over a limited range of system scale, the fact that astrophysics is ultimately scale-dependent implies the need for a more flexible approach to describing population statistics over a wide dynamic range.  For such applications, we introduce and implement a class of \textsc{Kernel-Localized Linear Regression (KLLR)} models.  \textsc{KLLR} is a natural extension to the commonly-used linear models that allows the parameters of the linear model -- normalization, slope, and covariance matrix -- to be scale-dependent. \textsc{KLLR} performs inference in two steps: (1) it estimates the mean relation between a set of independent variables and a dependent variable and; (2) it estimates the conditional covariance of the dependent variables given a set of independent variables. We demonstrate the model's performance in a simulated setting and showcase an application of the proposed model in analyzing the baryonic content of dark matter halos. As a part of this work, we publicly release a Python implementation of the \textsc{KLLR} method.
\end{abstract}


\section{Introduction} 

Multivariate linear regression of the simple least-squares kind has  been a canonical method used to characterize the scaling relations among a set of physical properties or other system variables \citep{Isobe1990LinRegr, Kelly:2007}.  Such linear models are extensively used in astronomy to model observational data, analyze simulated data, and compare empirical data with theoretical models. 
Their utility, however, is limited by the 
relative simplicity of the assumptions underlying the methods.
The key assumptions of the least-square linear model are:  \emph{homoscedasticity}, there exists a common global variance that is independent of the input variables;  \emph{scale independence}, the slopes and normalization are independent of the relevant scale in the problem, and; \emph{normality}, the noise is Normally distributed. These strong assumptions have constrained the scope of these models in enabling new discoveries.
With larger data collection facilities and faster computing machines, the volume, dimension, and complexity of empirical and simulated data are rapidly expanding; and as a result the traditional linear models are becoming a limiting factor in extracting accurate statistical summaries of increasingly dense data collections.

Non-linearity induced by scale-dependent physics between system-level properties provides extra information that can be exploited to gain insights into the governing dynamics of astronomical systems. 
Non-linear, scale-dependent trends both in the mean and variance of a population are common in many astronomical systems such as cluster- and group-size halos \citep[{\sl e.g.},][]{Farahi:2018,Anbajagane2021sigma_dm, Anbajagane2021vel_bias}, galaxies \citep[{\sl e.g.},][]{Cappellari2013, Mowla:2019, Eadie:2021, Anbajagane2021sigma_dm}, globular clusters \citep[{\sl e.g.},][]{Fahrion:2020}, among others.
Thus, there is a need for analysis tools that can discover and measure these non-linear signals both in the mean relation and variance about the mean relation.

One technique, generalized linear models \citep[{\sl e.g.},][]{DeSouza:2015,Elliott:2015}, allows non-linearity by introducing a \emph{link function} $g(.)$ that relates the mean of a linear relation to the expected value of the response variable. The scale dependence can be hard coded into the link function $g(.)$, the shape of which should be known prior to inference. These models are among fully specified models and their degree of freedom is comparable to those of simple linear models. Despite their simplicity and interpretability, the fact that the \emph{link function} needs to be parameterized restricts their applications to a subset of problems where the relation between the independent and response variables is known.

Parameter-free models, such as Bayesian additive regression trees \citep[][]{Hill2020bayesian}, Gaussian processes \citep[][]{Alvarez:2011}, ensemble tree models \citep[][]{chen2016xgboost}, and neural networks \citep[][]{rumelhart1986learning}, have emerged as an alternative to fully specified models \citep[{\sl e.g.},][]{Green2019,deSouza2021,Machado:2021,Stiskalek:2022,Ntampaka:2022}. Despite unmatched success of these models in making accurate predictions, their applications to physical sciences, including astronomy, can be limited by their interpretability \citep[][]{Ntampaka:2022}.
It is also an arduous task to incorporate extra physical assumptions into these models \citep{narasimhan2018learning,Seo:2021} and learn the variance about the mean relation \citep[][]{liitiainen2009residual}.  
To make the most of these models, the scientific community continues developing innovative solutions to these outstanding drawbacks \citep[{\sl e.g.},][]{Lundberg2017unified,Ponte:2017,devroye2018nearest,Ntampaka:2019}.

Another class of non-linear models, that blend properties of generalized linear models with additive models, is the generalized additive models \citep[][]{hastie2017generalized}. These models keep the interpretability of fully specified models and combine it with the flexibility of parameter-free models. 
A class of closely related models that is particularly suitable for scale-dependent problems is the method of local polynomial regression \citep[LOESS and LOWESS][]{cleveland1979robust,cleveland1996smoothing,takezawa2005introduction}.
These models are globally unspecified, which enables discovering new physics, while locally reducing to polynomial models, which makes them interpretable.

In this work, we implement a variation of the local polynomial regression model that is designed to explicitly capture the scale-dependent trends in the relations between observables. Specifically, we relax (1) homoscedasticity and (2) scale independence assumptions that are behind the least-square linear model, while keeping the model interpretable and computationally tractable by allowing the normalization, slope, and variance to vary with a chosen \emph{scale} of the problem. 
Our objective is to estimate a mean, $\mathbb{E}(\boldY \mid \boldX)$, and covariance, ${\rm Cov}(\boldY \mid \boldX)$, by assuming a locally linear, but globally non-linear, relation between variables $\boldX$ and $\boldY$. By \textit{locally linear}, we mean that near a chosen independent scale (designated $x_1$ below), the dependent variables can be approximated as linear in the mean with some local covariance about that mean.  Both the linear parameters and covariance are allowed to vary with the scale parameter, the choice of which is defined by the physics of the problem. 

To this end, we propose the \textsc{Kernel Localized Linear Regression} (\textsc{KLLR}) model class that allows the parameters of the least-square linear model to become scale-dependent. By relaxing the assumption of scale independence, this model class provides a more nuanced, but still easily interpretable, description of population statistics that is appropriate for large samples that have broad dynamical range and contain non-linear trends.

Accompanying this paper's publication, we release the open-source software \textsc{Kernel Localized Linear Regression} (\textsc{KLLR}, \href{https://github.com/afarahi/KLLR}{\faGithub} )\footnote{\url{https://github.com/afarahi/KLLR}} at the disposal of the community. The \textsc{KLLR} is a Python package that is indexed by Python Package Index (PyPI) and can be installed through \texttt{pip install kllr}.

While this package is developed and implemented with applications in data analysis of astronomical data sets in mind, its applications are not limited to the astronomy domain alone.  The analysis of systems covering a wide range of scale and having multiple, interrelated properties is relevant to a variety of subjects, including econometrics, population ecology, and neuroscience. In this implementation, we assume that the independent and dependent variables are noiseless. However, the code allows for measurement uncertainties in the dependent variables with the limitations discussed in \S~\ref{sec:limitations}.

In \S \ref{sec:formulation}, we begin with the problem setup and introduce the notation. In \S \ref{sec:software}, we discuss some of the key features of \textsc{KLLR} package and its dependencies. In \S \ref{sec:application}, we demonstrate an application of the proposed model for astronomical data analysis.  Finally, we conclude this work in \S \ref{sec:conclusion}.

\section{Problem Setup}  \label{sec:formulation}

In this section we introduce the model behind \textsc{KLLR}. We start with introducing the notation and setting up the regression problem. We then introduce the case of measurement noise in the data and illustrate how it can be handled within this setting. Finally, we discuss the limitations of this method and practical considerations in its use.

\subsection{Notation}

We denote the independent variable by $\boldX$ and the dependent variable\footnote{In statistics literature, $\boldX$ and $\boldY$ are often known as the ``covariate'' and the ``response'' variables, respectively.} by $\boldY$.  $\boldX$ and $\boldY$ are $d$-dimension and $m$-dimension vectors, respectively. In the first part we assume both $\boldX$ and $\boldY$ are perfectly measured; but in the next section we allow for uncertainty on the dependent variable as well. Our goal is to find the relation between $\boldX$ and $\boldY$. The vector $\boldX$ has a specific structure; the first element of $\boldX$ is the scale variable $\mu \equiv \boldX_1$ and the rest are ordinary regression variables.

Throughout this work, $\langle \cdots \rangle$ is employed to denote the expectation value. $\langle \boldY\,|\,\boldX \rangle$ denotes the expectation value of $\boldY$ given $\boldX$. $\given{\boldY}{\boldX}$ denotes the random variable $\boldY$ conditioned on $\boldX$. $i$ is the index over data points. $\langle \boldY_i \rangle$ is a shorthand for  $\langle \boldY_i\,|\,\boldX_i \rangle$. The property vector of an astronomical system is a random variable

\begin{equation} \label{eq:model}
    \given{\boldY_i}{\boldX_i} = \langle \boldY\,|\,\boldX_i \rangle + \boldsymbol{\epsilon}(\boldX_i),
\end{equation}

where $\boldsymbol{\epsilon}_i \equiv \boldsymbol{\epsilon}(\boldX_i)$ is a random variable described with a multivariate normal distribution with mean zero. $\boldsymbol{\epsilon}$ defines the intrinsic randomness in the conditional property $\boldY$ given $\boldX$. In general, the noise can be a function of $\boldX$. We limit the model class to the class of functions that the noise variable is only a function of the scale variable $\mu$.

For now, we will assume $\boldY$ and $\boldX$ are measured perfectly, and their measurement noise is negligible with respect to the variance of the population. However, later in \S \ref{sec:measurment_noise}, we allow for $\boldY$ to be uncertain, and we introduce the relevant notation here.
If the measured quantities are noisy then the noise-affected measurement and the corresponding noise covariance matrix are denoted by $\hatboldY$ and $\boldSigmaErr$, respectively. We further assume that $\boldSigmaErr$ is diagonal. If a noisy version of $\boldY_i$ is measured, this measurement is another random variable 
\begin{equation}
    \given{\hatboldY_i}{\boldY_i} = \boldY_i + \boldsymbol{\epsilon}_{i, \rm err},
\end{equation}
where $\boldsymbol{\epsilon}_{i, \rm err}$ is the measurement noise vector, which is a random variable drawn from a multivariate normal distribution with mean zero.

While we allow for uncertainty on the dependent variable, the user should consider that this model is not designed for noisy data. If the noise level is smaller than the intrinsic scatter and the sample size is large enough, the model works fine. Otherwise the user needs to establish the robustness of estimated quantities independently for their application needs (see \S \ref{sec:limitations} for a discussion on this).   

Because of the multivariate normal assumption in Equation~\eqref{eq:model}, our model can be fully described with two quantities: (1) the expected conditional property $\langle \boldY\,|\,\boldX \rangle$ -- the mean trend -- and (2) the conditional covariance $\boldSigma \equiv {\rm Cov}(\boldY\,|\,\boldX)$. Our goal is to estimate these two quantities in two steps by employing a scale-dependent but locally linear model.

\begin{table*}
\caption{Notations. {\bf Top block.} Data specification. {\bf Second block.} Model specification. {\bf Third block.} Kernel function setup. {\bf Fourth block.} Index notation. {\bf Bottom block.} Data size and dimension.} \label{tab:notation}
	\begin{center}
		\tabcolsep=0.8mm
		\begin{tabular}{ | l | l | l | }
            \hline
            Parameter & Explanation & Category \\
            \hline
            $\boldX$ & Independent variables vector. & Input variable.\\
            $\mu$ & Scale variable, $\mu=\boldX_1$. & Input variable.\\
            $\boldY$ & Dependent property vector. & Random variable. \\
            $\hatboldY$ & Dependent property vector, if $\boldSigmaErr \neq 0$. & Random variable. \\
            $\boldSigmaErr$ & Measurement error matrix (assumed diagonal). & Constant. \\
            \hline
            $\alpha$ &  Scale-dependent normalization. & Model Parameter. \\
            $\boldb$ & $d$-dimensional scale-dependent slope. & Model Parameter. \\
            $\boldSigma$ & Conditional covariance matrix, conditioned on $\boldX$. & Model Parameter.  \\
            \hline
            $k(.,.)$ & Kernel function, we assume a Gaussian kernel. & Function.\\
            $l$ & Smoothing scale in the Gaussian kernel. & Hyperparameter. \\
            \hline
            $i$ & Index over data points. & Index.\\ 
            $j$ & Index over the vector of dependent variable. & Index. \\ 
            \hline
            $n$ & Number of data points. & --\\ 
            $d$ & Dimension of vector $\boldX$. & -- \\ 
            $m$ & Dimension of vector $\boldY$ and $\hatboldY$. & -- \\
            \hline
    \end{tabular}
	\end{center}
\end{table*}

\subsection{Inferring the Mean Relation} \label{sec:regression}

As a key feature, the \textsc{KLLR} method employs a scale-dependent regression model to estimate the conditional expected properties. 
The relevant scale parameter of the problem is specified with the first variable of vector $\boldX$ which will be denoted with $\mu$. Suppose a scale-dependent linear model
\begin{equation} \label{eq:scaling}
   \langle y_j\,|\,\boldX \rangle = \alpha_j(\mu) + \boldb_j(\mu) \cdot \boldX,
\end{equation}
where $\alpha_j(\mu)$ and $\boldb_j(\mu)$ are the scale-dependent   normalization and slope.  The dot product, $\boldb_j(\mu) \cdot \boldX$, is relevant for cases in which a $d$-dimensional independent variable is used, but the single scale parameter element, $\mu$, determines the localization.  

In this section, $\langle y_j \,|\,\boldX \rangle$ is fitted for each observable $j$ independently. Thus, without loss of generality, we suppress the index $j$ in the rest of this section.
Following \citet{Farahi:2018},  $\alpha(\mu)$ and $\boldb(\mu)$ are estimated at fixed $\mu$ by minimizing the square weighted error
\begin{equation}
\epsilon^2(\mu) = \min_{\boldb, \alpha} \sum_{i=1}^n \ w_i^2(\mu) \ \left[y_{i} - \boldb(\mu) \cdot \boldX_i - \alpha(\mu)\right]^2,
\end{equation}
where the sum $i$ is over all data points and $w_i(\mu)$ is the local weight centered on a chosen scale $\mu$. At fixed $\mu$, the weights are 
\begin{equation}
w_i(\mu) = k(\mu_{i}, \mu),
\end{equation}
where $k(\mu_{i}, \mu)$ is a kernel function. Sweeping through steps in $\mu$ produces the scale-dependent slope and normalization.

A top-hat kernel is equivalent to a binning strategy that is typically employed in the astronomy literature. While the software allows the user to chose between a top-hat and a Gaussian kernel, we advocate for a Gaussian kernel which has a smooth form given by
\begin{equation}
w_i(\mu) \propto \exp \left[-\frac{(\mu - \mu_i)^2}{2l^2}\right],
\end{equation}
where $l$ is the width of the Gaussian kernel that specifies the smoothness of the inferred quantities. While there is no unique, optimal choice for $l$, there are strategies that can guide the user to an acceptable value for the smoothing scale; see the discussion in \S \ref{sec:limitations} for more details.

\subsection{Inferring the covariance matrix} \label{sec:cov_inference}

Our second aim is to infer the conditional covariance matrix of two properties $\boldY_j$ and $\boldY_{j^{\prime}}$ at fixed $\mu$. 
The key assumption is that the covariance matrix is only a function of the scale variable $\mu$ and no other independent variables. We do not parametrize the covariance matrix; instead, as before, we estimate it at fixed $\mu$. 
Similar to the slope and normalization estimation, we assume that the covariance matrix is a slowly varying function of $\mu$ with respect to the smoothing scale factor $l$.

The data consist of a vector of observed properties denoted by  $\boldY$, a random vector of $m$-dimension, at a fixed $\mu$. We assume that the conditional distribution of $\boldY$ given $\boldX$ is described by a multivariate Gaussian distribution,
\begin{equation} \label{eq:cov_data_model}
   \given{\boldY}{\expymu, \boldSigma(\mu), \mu} \sim  \mathcal{N}(\expymu, \boldSigma({\mu})).
\end{equation}
The mean and the covariance of this conditional distribution are denoted with an $m$-dimensional vector $\expymu$ and an $m \times m$ matrix $\boldSigma(\mu)$, respectively. The mean vector is
\begin{equation}
    \langle y_{j} \rangle_{\mu} = \langle y_j \,|\,\boldX \rangle = \alpha_j(\mu) + \boldb_j(\mu) \cdot \boldX.
\end{equation}
that is estimated in the previous section. 
The covariance matrix $\boldSigma({\mu})$ can be specified with three independent parameters, the scatters of two variables $\sigma_{j}(\mu)$ and $\sigma_{j^\prime}(\mu)$ and then a correlation matrix $r_{j,j^{\prime}}(\mu)$ \citep{barnard2000modeling}. To estimate these quantities we first define a residual vector
\begin{equation}
    \delta y_{i,j} =  y_{i,j} - \langle y_{i,j} \rangle_{\mu}.
\end{equation}

The property covariance in our weighting scheme can be readily estimated. We use an unbiased weighted estimator of the covariance matrix $\Sigma(\mu)$ \citep{GNU:2009}, 
\begin{equation} \label{eq:r-estimator}
\Sigma_{j, j^{\prime}}(\mu) = A \sum\limits_{i=1}^{n}w_{i}(\mu) ~ \delta y_{i,j} ~ \delta y_{i,j^{\prime}},
\end{equation}
where 
\begin{equation}
    A = {\frac {\sum\limits_{i=1}^{n}w_{i}(\mu)}{\left(\sum\limits _{i=1}^{n}w_{i}(\mu)\right)^{2}-\sum\limits _{i=1}^{n}w_{i}^{2}(\mu)}}
\end{equation}

Using the above estimator, we compute the parameters of interest -- the property scatters and the correlation coefficients. An estimator of scatter is 
\begin{equation}
    \sigma_{j}(\mu) = \sqrt{\Sigma_{j, j}(\mu)}
\end{equation}
and the correlation matrix can be estimated with 
\begin{equation}
    r_{j, j^{\prime}}(\mu) = \frac{\Sigma_{j, j^{\prime}}(\mu)}{\sqrt{\Sigma_{j, j}(\mu) ~ \Sigma_{j^{\prime}, j^{\prime}}(\mu)}}
\end{equation}
which reduces to $r_{j,j^{\prime}}=1$ if $j = j^{\prime}$.

\subsection{Model diagnosis}

While the assumptions made in this work are weak in comparison to those typically made in the astronomy literature, it is still advised that the user check whether the model assumptions hold for their input data. The \textsc{KLLR} software is equipped with diagnosis test methods that allows the user to check the validity of the model assumptions. We describe these tests below.

The two key assumptions of the KLLR model are (1) the conditional likelihood of the dependent variable $\boldY$ can be described with a multivariate normal distribution, and (2) the model parameters (normalization, slope, and covariance) are only a function of scale variable but not the rest of independent variables in $\boldX$.

One way to validate the multivariate normal assumption is to compute the higher moments of the residuals about the mean relation such as skewness, $\boldsymbol{\gamma}$, and kurtosis, $\boldsymbol{\kappa}$. Skewness is given by
\begin{equation} \label{eqn:Skew_definition}
    \boldsymbol{\gamma} = \frac{\langle(\boldY - \expymu)^3\rangle}{\langle (\boldY - \expymu)^2\rangle^{3/2}} 
\end{equation}
and kurtosis is given by
\begin{equation} \label{eqn:Kurosis_definition}
    \boldsymbol{\kappa} = \frac{\langle(\boldY - \expymu)^4\rangle}{\langle(\boldY - \expymu)^2\rangle^2}
\end{equation}
where $\langle(\boldY - \expymu)^2\rangle$ is more familiarly known as the variance $\sigma^2$.

Besides estimating higher order moments, the quantile-quantile (Q-Q) plot is another tool for evaluating the normality assumption. 
It is a visualization technique for determining if a population sample comes from an assumed distribution, here a normal distribution \citep[see Figure 4 in][]{Farahi:2018}.  
In the astronomy literature, there are a few works, such as that by \citet{Mantz:2008}, that employed the Q-Q plot and illustrated the consistency of the model assumption and data distribution.

The \textsc{KLLR} method focuses on  the first approach and is equipped with an estimator of skewness and kurtosis, as well as moments of arbitrary order. Both are computed as being globally scale-dependent but locally linear and in a manner completely analogous to the scatter measurement. The expectation values $\langle(\boldY - \expymu)^m\rangle$ in Equation~\eqref{eqn:Skew_definition} and Equation~\eqref{eqn:Kurosis_definition} are computed as weighted averages, where the weights are defined as before. In the limit where the normality assumption is exact, we obtain $\gamma = 0$ and $\kappa = 3$. However, $\gamma = 0$ and $\kappa = 3$ does not necessarily imply a Gaussian distribution as higher order moments might deviate from a Gaussian expectation.

In most applications, we are interested in distributions that are ``close enough'' to a Gaussian distribution. But how to quantify whether a model is close enough or not depends on the application.
The science requirements of a survey determine what range of non-Gaussianity is acceptable in an analysis.  
For instance, non-Gaussian scatter induces bias in the estimated halo mass function that can be quantified using perturbation theory \citep{Shaw2010}. Given the accuracy required by a survey one can put a bound on the acceptable levels of $\gamma$ and $\kappa$ and then evaluate if the measured non-Gaussinity is within the acceptable range or not.

Next, the $\boldX$-independence\footnote{By $\boldX$-independence, we mean the dependence on independent variables \textit{other} than the first element of $\boldX$, which is the scale variable $\mu$.} of the model parameters can be determined by binning the data in $\boldX$ and independently estimating the model parameters for each bin. 
The binning has to be done manually by the user. For instance, the user may want to split their sample on $\boldX_2$ and check if they get similar scaling parameters for each subsample. Since there are many ways of subdividing the sample in a high-dimensional setting -- this process is not automated and the user has to manually perform this diagnosis test, if desired. The user can easily bin their data, run \textsc{KLLR} for each subset and check the independence assumption.

\subsection{Parameter estimation in uncertain setting} \label{sec:measurment_noise}

Estimation in uncertain settings, where only noisy measurements are available, is rather common in astronomy \citep{Hogg:2010,Andreon:2013}. 
There are parametric models \citep[{\sl e.g.},][]{Kelly:2007,Sereno:2015,Mantz:2016,Sereno:2016} and non-parametric models \citep[{\sl e.g.},][]{POPE:2021} that are designed specifically to perform inference in uncertain settings.
The parametric models are limited by the parameterization imposed by each model. These models can be considered as a variation of the linear model that allows for inference in uncertain settings but their underlying model is the same. 
Non-parametric models, such as \textsc{PoPE} \citep{POPE:2021}, are comparable to \textsc{KLLR}, but can be computationally expensive for large sample sizes and require binning as well.
To broaden the applicability of the \textsc{KLLR} method, we extend the proposed estimators to handle measurement noise. 

In uncertain settings, we do not observe the actual values of vector $\boldY$, and instead observe values of $\hatboldY$ which are measured with measurement error $\boldSigmaErr$.
The measured quantity is assumed to be drawn from a multivariate Gaussian distribution, 
 \begin{equation}
    \given{\hatboldY}{\boldY} \sim  \mathcal{N}(\boldY, \boldSigmaErr).
 \end{equation}
Furthermore, $\boldSigmaErr$ is assumed to be diagonal so that there is no correlation between measurement errors of two quantities. 

In a heteroscedastic setting, $\boldSigmaErr$ varies with data; and each data point $i$ is generated by a multivariate Gaussian distribution,
\begin{equation}
    \given{\hatboldY_i}{\langle \boldY_{i} \rangle, \boldSigma} \sim  \mathcal{N}(\langle \boldY_{i} \rangle, \boldsymbol{\Sigma}_{\rm err, i} +\boldSigma),
\end{equation}
where $i$ is the index over the data and $\boldsymbol{\Sigma}_{\rm err, i}$ is the error covariance for data point $i$. We remind the reader that $\Sigma$ is the ``true'', intrinsic covariance of the data. The expected mean property $\langle \boldY_{i} \rangle$ is a function of independent variables $\boldX_i$ as defined in Equation~\eqref{eq:scaling}. 

Given a set of observations $\{\hatboldY_i, \boldX_i, \boldsymbol{\Sigma}_{\rm err, i}\}_{i=1:n}$, we want to estimate the normalization, slope, and the covariance matrix at fixed scale $\mu$. We can use the same estimators described in \S \ref{sec:regression} and \S \ref{sec:cov_inference} with the following modifications.

First, the weights shall be modified as
\begin{equation}
w_{i,j}(\mu) = k(\mu_{i}, \mu) / \sigma_{\rm  err,i,j},
\end{equation}
where $\sigma_{{\rm  err},i,j}$ is the measurement uncertainty of sample point $i$ and property $j$.
This model gives more weight to the data points with smaller measurement uncertainty as they provide more information compared to measurements that are highly uncertain. In a heterogeneous setting where the uncertainty for all measured quantities is the same, $\sigma_{\rm err}$ is just a normalization factor that can be ignored. The estimators for slope and normalization remain unbiased.

Estimating scatter requires additional correction due to the excess of observed variance because of the measurement noise. Scatter may be estimated by employing a weighted average of 
\begin{equation}
    \sigma^2_{j}(\mu) = A \sum_{i}^{n} w_{i,j} \left[ (\hatY_{i,j} - \langle y_{i,j} \rangle_{\mu})^2 - \sigma^2_{{\rm  err},i,j} \right]
\end{equation}
with $A$ defined as before and we omit the explicit $\mu$ dependence of $w$ for simplicity. Similarly, the correlation coefficient can be estimated with 
\begin{equation} \label{eq:corr_noisy_data}
    r_{j,j^{\prime}}(\mu) = \frac{ \sum_{i}^{n} \sqrt{w_{i,j}~w_{i,j^{\prime}}} \left[ (\hatY_{i,j} - \langle y_{i,j} \rangle_{\mu}) (\hatY_{i,j^{\prime}} - \langle y_{i,j^{\prime}} \rangle_{\mu}) - \sigma_{{\rm  err},i,j} \, \sigma_{{\rm  err},i,j^{\prime}} \right] } {\sqrt{ \sum_{i}^{n} w_{i,j} \left[ (\hatY_{i,j} - \langle y_{i,j} \rangle_{\mu})^2 - \sigma^2_{{\rm  err},i,j} \right] \sum_{i}^{n} w_{i,j^{\prime}} \left[ (\hatY_{i,j^{\prime}} - \langle y_{i,j^{\prime}} \rangle_{\mu})^2 - \sigma^2_{{\rm  err},i,j^{\prime}} \right] }}.
\end{equation}

where $\sum$ denotes a sum over objects indexed by $i$, and \textit{not} the covariance matrix $\Sigma_{jj\prime}$. We also note that the above estimators are biased and not strictly positive.
Due to these limitations, the estimated covariance (correlation and scatter) can lead to spurious conclusion. Thus, the \textsc{KLLR} method is not appropriate for estimating covariance matrix of noisy samples; and Equation~\eqref{eq:corr_noisy_data} is not implemented as a part of \textsc{KLLR} package. We discourage the use of this feature when data are noisy. 

We emphasize that the \textsc{KLLR} method is primarily designed to analyse data with negligible measurement uncertainty. Most importantly, it is a particularly strenuous task to incorporate uncertainty of the independent variable into an estimator. Inference through generative models in a Bayesian framework is more suitable for these settings \citep{Hogg:2010}.

\subsection{Limitations and extra considerations} \label{sec:limitations}

\noindent {\bf Inference in the Presence of Measurement Noise.} The applications of the modified estimators for noisy measurements are limited to settings where the average measurement noise is smaller than the intrinsic scatter.
This is appropriate specially for simulation data, where the measured quantities can be noisy due to sub-sampling, simulation resolution, or other reasons but their uncertainties do not dominate the signal.
Our model works for observational data in a limit of high signal-to-noise ratio (SNR) measurements.
In a low SNR regime, where the uncertainties dominate the signal, the \textsc{KLLR} model is limited and the user might want to consider other options in the market such as \textsc{PoPE} \citep{POPE:2021}, \textsc{LRGS} \citep{Mantz:2016}, \textsc{LIRA} \citep{Sereno:2016}, \textsc{LinMix} \citep{Kelly:2007}, among others.

\noindent {\bf Setting Kernel Width.} The estimated parameters asymptotically approach the true values in the limit of $l \rightarrow 0$ and $n \rightarrow\infty$, where $n$ is the sample size, and $l$, as a reminder, is the width of the (Gaussian) kernel.
In a finite sample setting, though, $l$ should be fine-tuned. Large (small) $l$ results in underfitting (overfitting) the model. 
The optimal smoothing scale depends on the problem, the sample size, and the goals of the inference task. So, $l$ should be chosen based on the range of data, the number of data points, and the smoothness of the free parameters with respect to the scale of the problem. All of these vary with application. Furthermore, there exists no broadly accepted approach to optimize $l$. Here, we provide some suggestions that might help the user to set $l$.

This hyper-parameter essentially controls the trade-off between bias and variance of the estimator. Increasing the smoothing scale reduces the variance of the estimator but leads to a larger bias. 
Physically speaking, increasing the smoothing scale washes out small-scale features, such as those induced by noise variance, but preserves large-scale trends. Thus the physics of the problem can be used to guide the choice of $l$. 
Ideally, $l$ should be smaller than the scale of features expected to be extracted from data and small enough that the estimator's variance does not hinder inference by inducing considerable uncertainty on the estimated quantities.

Let's expand this salient point. The smoothing scale determines the scales at which the inferred quantities do not vary significantly. For instance, if the smoothing scale is set to $0.2$\footnote{In the unit of the scale variable $\mu$.} then rapid changes in slope and normalization that are smaller than $0.2$ will be washed out. 
We suggest the user vary this hyper-parameter to the point that uncertainties do not dominate the results. It is advised that the user performs a sensitivity analysis to ensure that slight variations in this hyper-parameter do not change their conclusions and results. 

In many applications where we deal with large-volume data, the smoothing scale is not a limiting factor, and slight variations to $l$ do not change the results and final conclusions. However, performing such a sensitivity analysis would be reassuring, as the physics should not depend on non-physical hyper-parameters that control the statistical properties of the estimator.

\noindent {\bf Non-uniform Population Density.} Another challenge arises from the fact that the population number density of astronomical objects is typically not uniformly distributed with respect to the scale variable $\mu$. For instance, suppose $\mu$ is halo mass, where the number density of massive dark matter halos decays exponentially with linear halo mass or is polynomial with log-halo mass \citep{Evrard:2014}. 
In these situations, it might be appropriate to consider the smoothing scale $l$ that itself is a function of $\mu$ or equivalently the number density. Suppose
\begin{equation}\label{eqn:N_halos_effective}
    N_{\rm eff} = \sum_{i = 0}^N \ w_i(\mu, l).
\end{equation}
One approach would be setting $l$ in such a way to keep the effective sample size $N_{\rm eff}$ constant for all $\mu$. 
$N_{\rm eff}$ can be set a-priori and used to estimate the hyper-parameter $l$ as a function of the scale variable $\mu$. This functionality of a $\mu$-dependent kernel scale, $l(\mu)$, is included in the KLLR package.

\noindent {\bf Quantifying Confidence Intervals.} Point estimations can be misleading --- in our case, reporting a slope or normalization that deviates from a theoretical prediction is of limited use unless uncertainties on the measured quantities are reported.
We suggest utilizing resampling algorithms to quantify uncertainty on the model parameters. 
\textsc{KLLR} is equipped with an implementation of the bootstrap resampling algorithm and uses this algorithm to estimate the statistical uncertainty on each model parameter.

\section{Software Details and Dependencies}\label{sec:software}

\textsc{KLLR} is a Python package for multivariate regression analysis. It enables the user to perform multivariate regression analysis and generate informative visualizations. It is an implementation of the kernel weighted linear regression method described in \S \ref{sec:formulation}.
The visualization modules seamlessly fit the \textsc{KLLR} model to a set of data, estimate the uncertainties, and produce a set of data products, summary statistics, and diagnostic test visualizations.

When the dependent variable, $\boldY$, is one dimensional, the user can plot the best fit, the local slope, and standard deviation as a function of scale variable, $\mu$, which is defined by the user. It also provides a module that visualizes the distribution of normalized residuals in $\boldY$. The user is particularly encouraged to investigate the normalized residuals since the model assumes the conditional statistics of the data follow a multivariate normal distribution. If there is any evidence of strong skewness or fatter or narrower tail than expected from the normal distribution, the \textsc{KLLR} model might not be suitable for that problem. When the dependent variable, $\boldY$, is multi-dimensional, the user can generate -- on top of the previously-mentioned one-dimensional features -- a visualization of the conditional covariance and correlation matrix (see the examples provided with the code). 

Another important feature of \textsc{KLLR} is that the user can split their data set into non-overlapping subsets based on a third quantity. Then, it performs multivariate regression analysis for each subset independently and visualizes the summary statistics on the same plot. We illustrate below a diverse set of use cases of the \textsc{KLLR} method.

The main function, \texttt{kllr\_model(...).fit(...)} from class \texttt{kllr\_model(...)} takes a vector of observables $\{\boldX, \boldY, \sigma_{\rm err}\}$, a kernel function and its hyper-parameters and find the best fit.
Passing $\sigma_{\rm err}$ is optional and if it is not provided by the user the function assumes that the measurements are noiseless.
While \textsc{KLLR} allows and considers the measurement error in estimating the model parameters, it is not designed to deal with noisy measurements and should be used with extra care. Ideally, it should be used when the noise level is significantly smaller than the intrinsic scatter or if the noise \textit{is} higher, then it should be used as an initial exploratory analysis, to guide formulation of a more accurate model.
\textsc{KLLR} performs regression and reports the local normalization, slope, standard deviation, and higher moments at each point in $\mu$. The current implementation of \texttt{kllr\_model(...)} supports a uniform and a Gaussian kernel with width defined by the user. It employs a bootstrap resampling algorithm to estimate the uncertainties for each model parameter. \textsc{KLLR} is backed by a set of user-friendly and fast visualization tools so practitioners can seamlessly generate informative data summaries and visualizations.

The visualization modules take \texttt{Pandas.DataFrame} objects as input and the user defines which columns are $\boldY$, $\boldX$, $\mu$, and the split variable. The user may set the value of the smoothing scale, if the default option is not desirable.

\subsection{Software dependencies}

The \textsc{KLLR} package uses \textsc{NumPy} \citep{NumPy} and \textsc{Scikit-learn} library \citep{Scikitlearn}, and plotting modules use \textsc{Pandas} \citep{Pandas} data structure to perform \textsc{KLLR} and visualize through \textsc{Matplotlib} \citep{Matplotlib}.

\subsection{Similar software packages}

\textsc{LOWESS} and \textsc{LOESS} are implementation of locally-weighted polynomial regression proposed by \citet{cleveland1979robust} for univariate and multivariate data, respectively. An implementation of both algorithms exists in Python\footnote{\url{https://pypi.org/project/loess/}} and R\footnote{\url{https://stat.ethz.ch/R-manual/R-devel/library/stats/html/loess.html}}. \textsc{LOESS} and \textsc{LOWESS} employ a weighting scheme similar to \textsc{KLLR} to perform regression; however, the weighting scheme in \textsc{KLLR} has the advantage of a built-in physical interpretation.
Additionally, \textsc{KLLR} reports an estimate of the scale-dependent covariance and equipped with a set of diagnosis tools as opposed to the \textsc{LOESS} and \textsc{LOWESS} implementations. This scale-dependent covariance contains information about the physical processes that govern the dynamical evolution of a population's observable properties \citep[][]{Farahi:2019,Anbajagane:2020, Anbajagane2021sigma_dm}.

Population Profile Estimator \citep[\textsc{PoPE},][]{POPE:2021} is another closely related software package. \textsc{PoPE} is a Bayesian inference model that uses Gaussian Processes to perform a regression task similar to what is done here. The key difference between these two models is that \textsc{PoPE} is designed to model low signal-to-noise ratio data while \textsc{KLLR} can only handle high signal-to-noise ratio data. See \citet{POPE:2021} for a comparison between \textsc{PoPE} and \textsc{KLLR}.

\subsection{A simulated example}\label{sec:examples}

In this example, we generate a simulated data set and illustrate the performance of the \textsc{KLLR} method. We assume a one-dimensional independent variable and a three-dimensional dependent variable. The mean relations are
\begin{align}\label{eq:modelmean}
    \langle y_1 \mid x \rangle &= -2x^2 + x \,, \nonumber\\
    \langle y_2 \mid x \rangle &= 2x^3 - x \,,  \\
    \langle y_3 \mid x \rangle &= -2 x - 2\,, \nonumber
\end{align}
The scaling of $y_1$ and $y_2$ are non-linear with respect to $x$ while $y_3$ is linear. The covariance matrix is
\begin{equation}\label{eq:modelcovar}
\Sigma = 
\begin{pmatrix}
\sigma_1^2 & r_{12} \sigma_1 \sigma_2  & r_{13} \sigma_1 \sigma_3  \\
r_{12} \sigma_1 \sigma_2  & \sigma_2^2  & r_{23} \sigma_2 \sigma_3 \\
r_{13} \sigma_1 \sigma_3  & r_{23} \sigma_2 \sigma_3 & \sigma_3^2  
\end{pmatrix},
\end{equation}
where
\begin{align}\label{eq:modelcovarparams}
    \sigma_{1} &= 0.5x^2 + 0.2 \,, \,\,\,\,\,
    \sigma_{2} = -0.1x^2 + 0.5  \,,\,\,\,\,\,  
    \sigma_{3} = 0.3\,, \nonumber\\
    r_{12} &= x^2 - 0.5\,,\,\,\,\,\, r_{13} = -x^2 + 0.5\,,\,\,\,\,\, {\rm and} \,\, r_{23} = 0 \,. 
\end{align}

For the independent variable, $x$, we draw 5,000 points uniformly sampled from $x \in [-1.2, 1.2]$. We then fit a \textsc{KLLR} model to this data and ask how well the model can recover the scaling parameters and covariance matrix.
We use a Gaussian kernel with a smoothing scale of $l=0.1$. The results are insensitive to the small changes in $l$. 
Like many potential applications in astronomy and beyond, this system has simple, smooth behaviors for which the impact of hyper-parameter tuning on the final results is negligible. To avoid edge effect, we only show the results for $x \in [-1, 1]$.

\begin{figure*}
     \centering
     \includegraphics[width=0.44\textwidth]{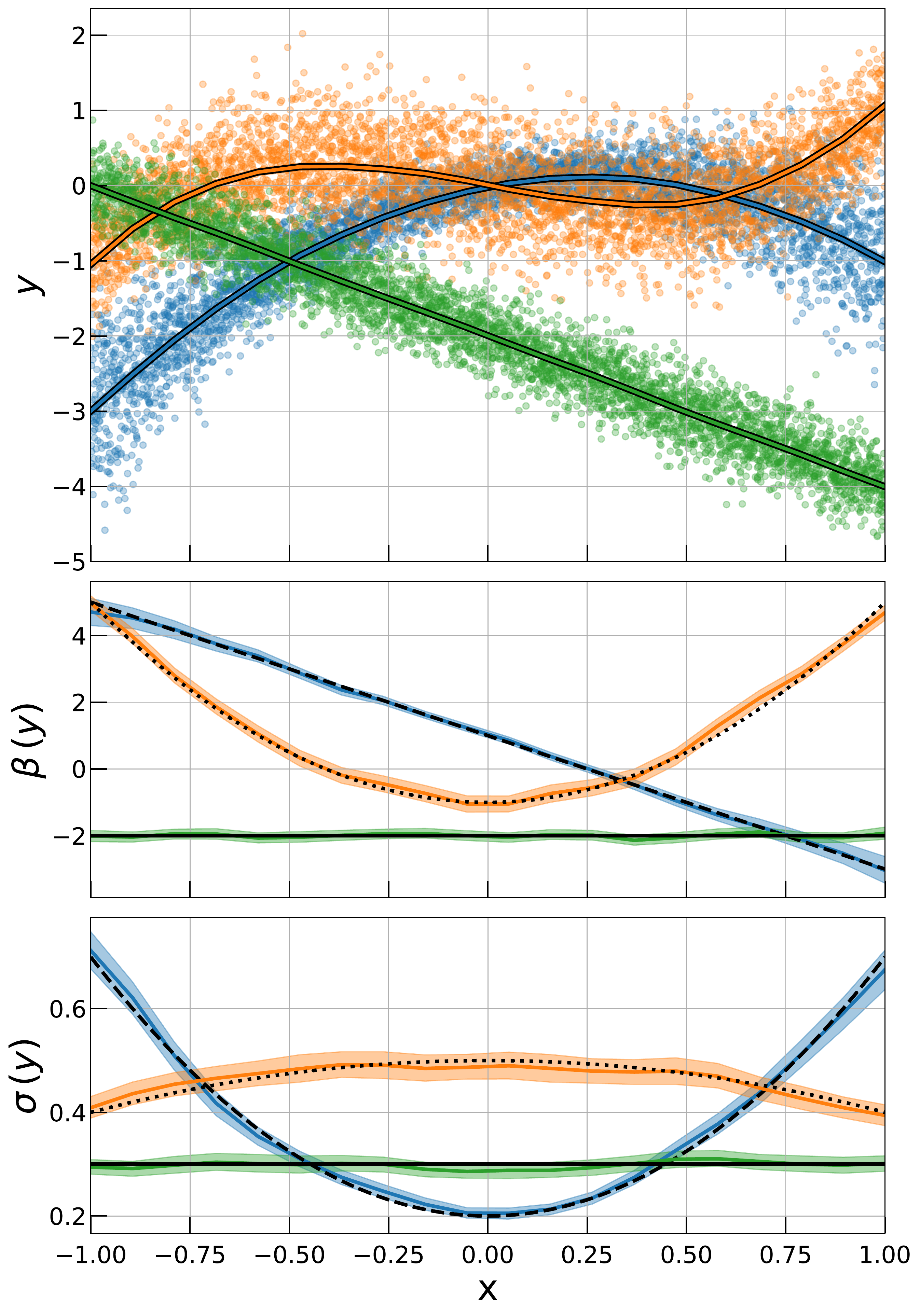} 
     \includegraphics[width=0.55\textwidth]{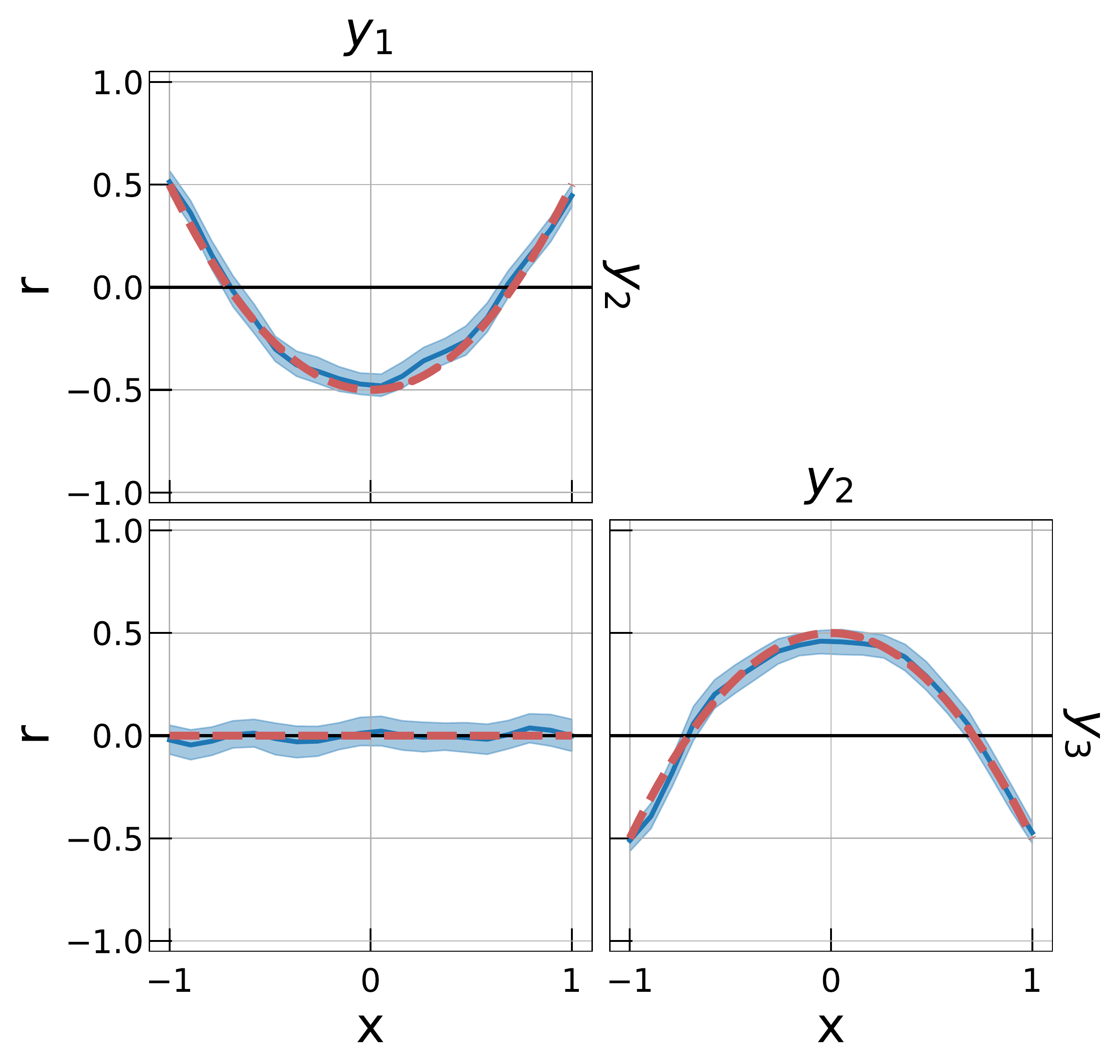} 
\caption{Illustration of the \textsc{KLLR} method using random samples of 5,000 points following three mean relations, Equation~\eqref{eq:modelmean}, with Gaussian covariance, Equations~\eqref{eq:modelcovar} and \eqref{eq:modelcovarparams}. {\bf Top Left Panel.} The data points and \textsc{KLLR} mean fits (solid lines), with each relation shown in a separate color.   {\bf Middle and Bottom Left  Panels.} The \textsc{KLLR} estimated slope and scatter as a function of the scale parameter, x. The black lines are the input parameters. Shaded regions are $95\%$ confidence intervals estimated using 1000 bootstrap realizations. Made using the \texttt{Plot\_Fit\_Summary()} function. {\bf Right Panel.} Recovered \textsc{KLLR} estimates of the scale-dependent correlation matrix,  Equation~\eqref{eq:modelcovarparams}. The red dashed-lines are the input correlations. Made using the \texttt{Plot\_Cov\_Corr\_Matrix()} function. }  
     \label{fig:fit_parameters_simulation}
\end{figure*}

Figure \ref{fig:fit_parameters_simulation} shows the extracted \textsc{KLLR} model parameters. The top left panel is the \textsc{KLLR} fit (solid lines) for $\{y_1, \boldX\}$ (blue line), $\{y_2, \boldX\}$ (orange line), and $\{y_3, \boldX\}$ (green line). The middle and bottom left panels shows the actual scale-dependent fit parameters. As expected the \textit{slope} for $y_2$ runs linearly with the scale variable $x$ while the slope of $y_3$ is scale independent and the slope of $y_3$ is a non-linear function of $x$. The scatter for two models is scale dependent and consistent with the diagonal elements of the input covariance matrix. The shaded regions are $95\%$ confidence intervals that are estimated using 1000 bootstrap data-realizations.  
The right panel presents the estimated correlation matrix as a function of the scale variable. \textsc{KLLR} recovers the scale-dependent input correlations. 

\begin{figure}
     \centering
     \includegraphics[width=0.52\textwidth]{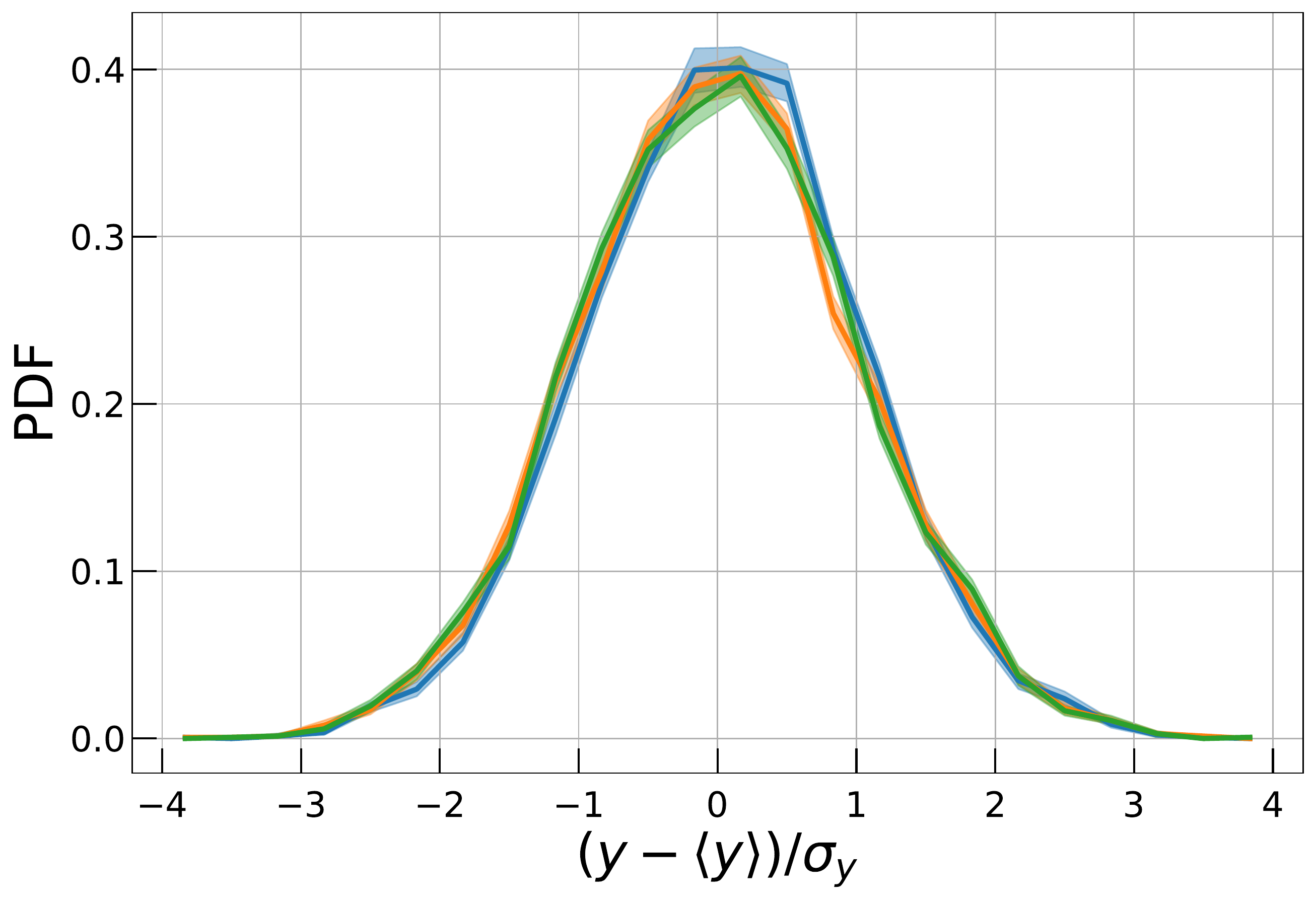} 
     \includegraphics[width=0.46\textwidth]{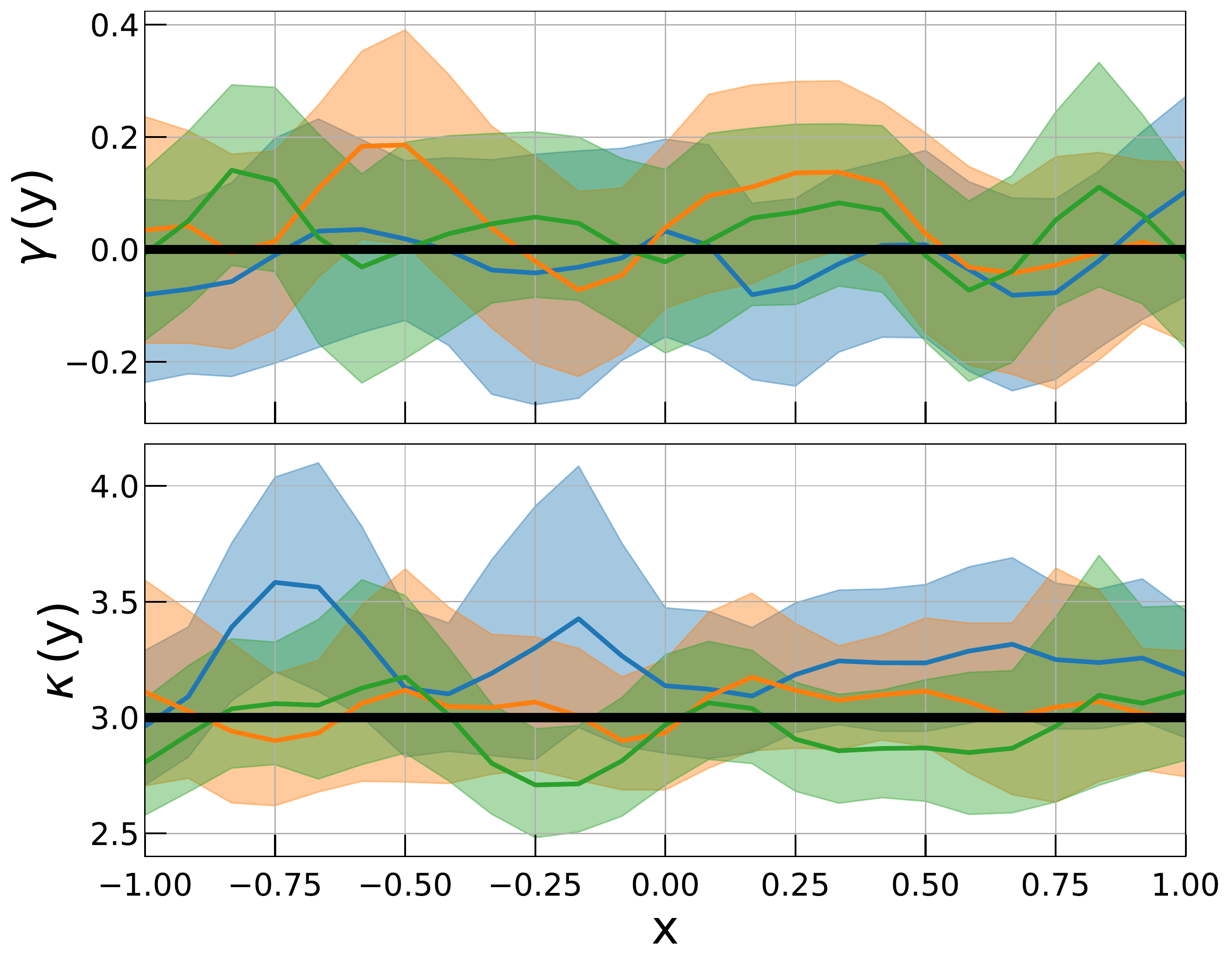} 
\caption{{\bf Left Panel.} Frequency distribution of residuals of the data points from the mean, normalized by the standard deviation, for model diagnosis. Made using the \texttt{Plot\_Residual()} function. {\bf Right Panel.} The third and fourth moment of each distribution (skewness and excess kurtosis) as a function of scale $x$. These higher order statistics are consistent with the assumed Normal distribution. Made using the \texttt{Plot\_Higher\_Moments()} function.}   
     \label{fig:diag_tests}
\end{figure}

Finally, Figure~\ref{fig:diag_tests} presents the result of our recommended diagnostic tests. The left panel shows the residuals in all three dependent variables; and the right panel shows the estimated skewness and kurtosis of the distributions as a function of the scale parameter, $x$. Using these diagnosis test, we confirm that the residuals follow the normal distribution. 

This illustrates the input model can be recovered with the \textsc{KLLR} method; and the output is easily interpretable. This example is provided with the package so that the user can both replicate the results and see how the code works. Now, we show some applications of this model to astronomical data analysis. 

\section{An application to astronomical data analysis} \label{sec:application}

\begin{figure}
    \centering
    \includegraphics[width = 0.49\textwidth]{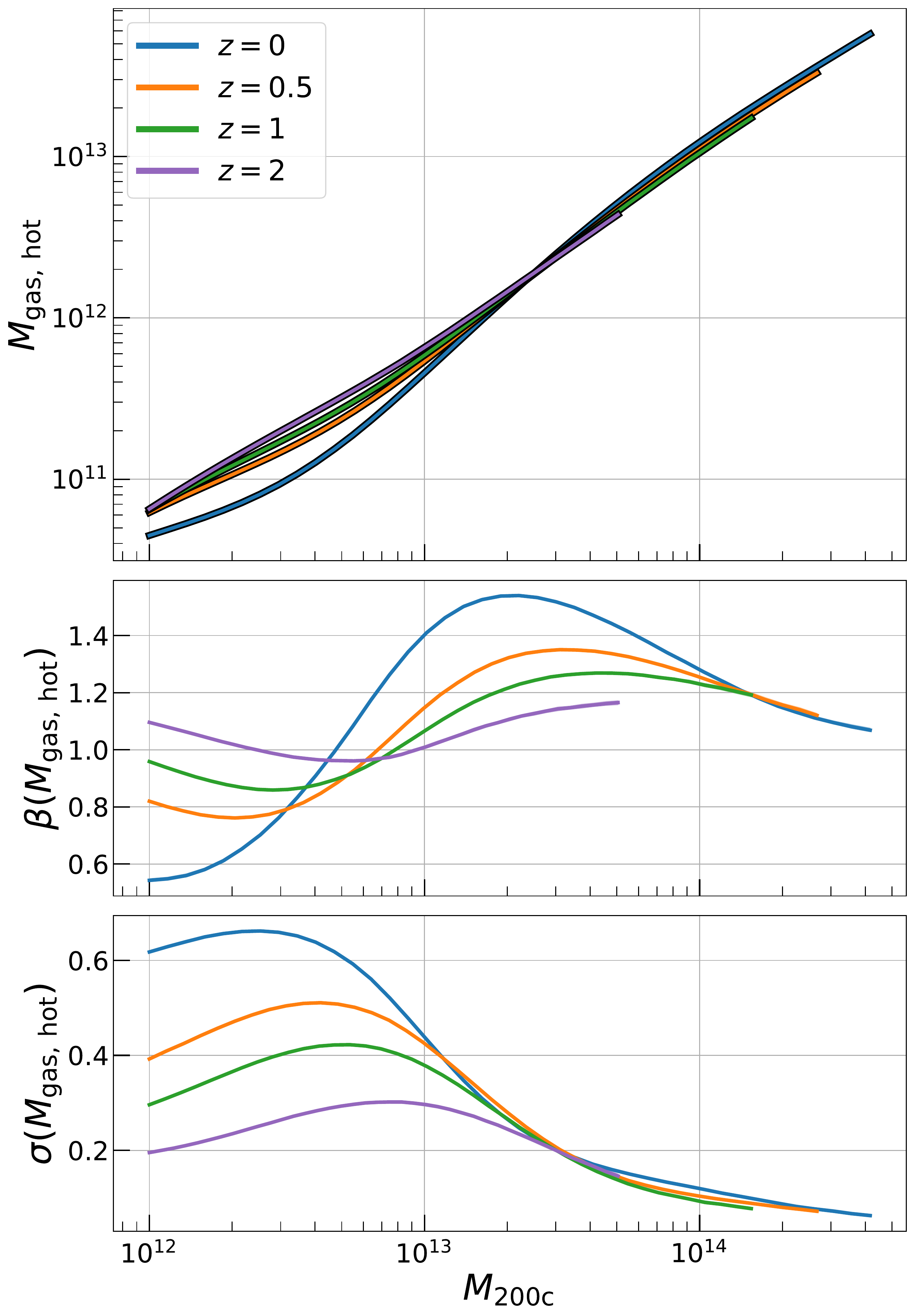}
    \includegraphics[width = 0.49\textwidth]{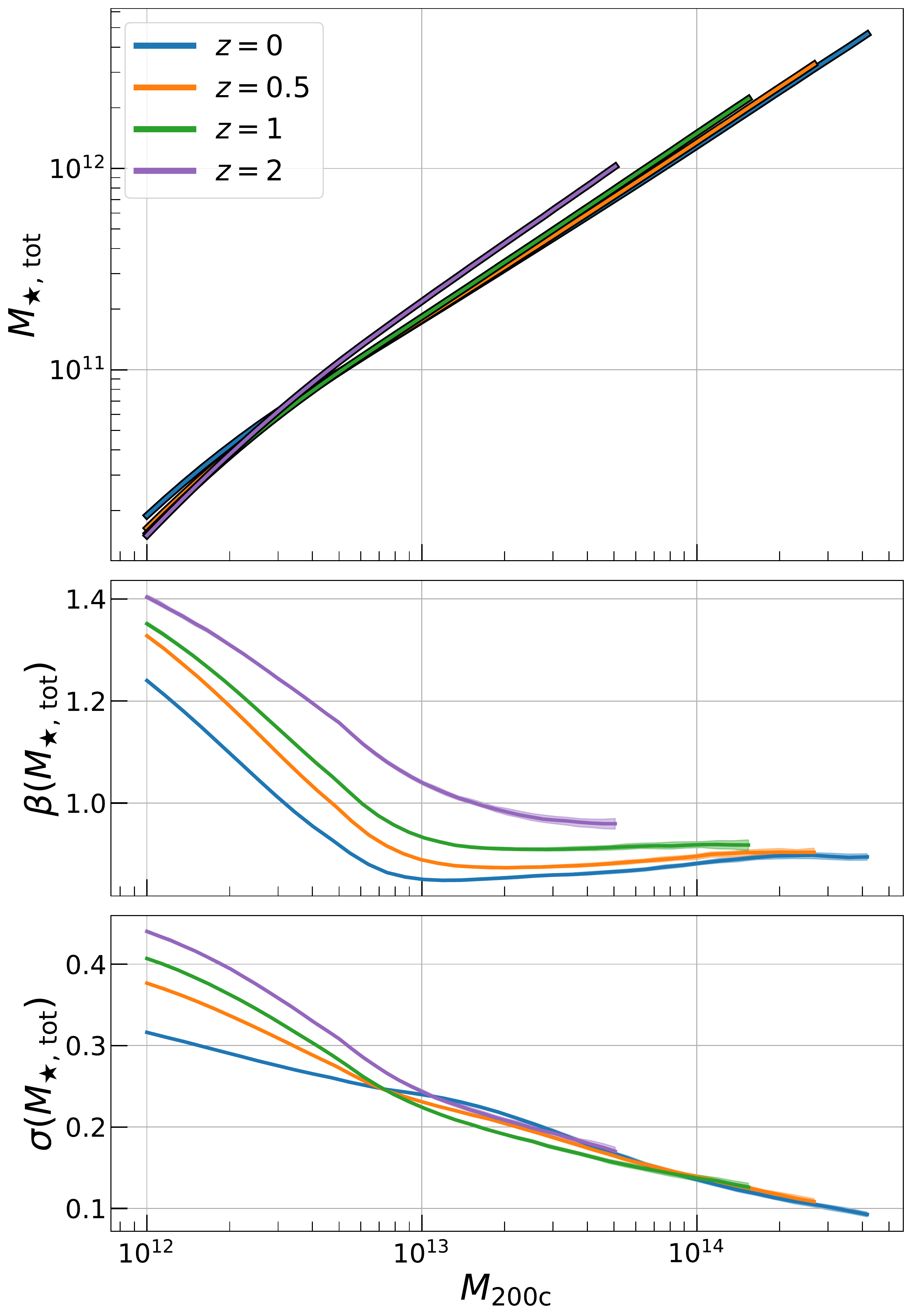}
    \caption{Local power-law scaling relation parameters returned by KLLR ---  normalization (top), slope (middle) and natural logarithmic scatter (bottom) --- for two halo properties, hot gas mass (left column) and total stellar mass (right) in the $z=0$ halo sample of the Illustris TNG300 simulation. Colors show population behaviors at different redshifts, indicated in the legend. Both relations feature time- and mass-dependent behaviors in all model parameters. The scatter, shown in natural log to facilitate interpretation as a fractional scatter, declines to below $10\%$ in both stellar and hot gas masses at the highest halo masses and is nearly redshift-independent.  Made using the \texttt{Plot\_Fit\_Summary()} function.}
    \label{fig:TNG_Fits}
\end{figure}

\begin{figure}
    \centering
    \includegraphics[width = 0.65\textwidth]{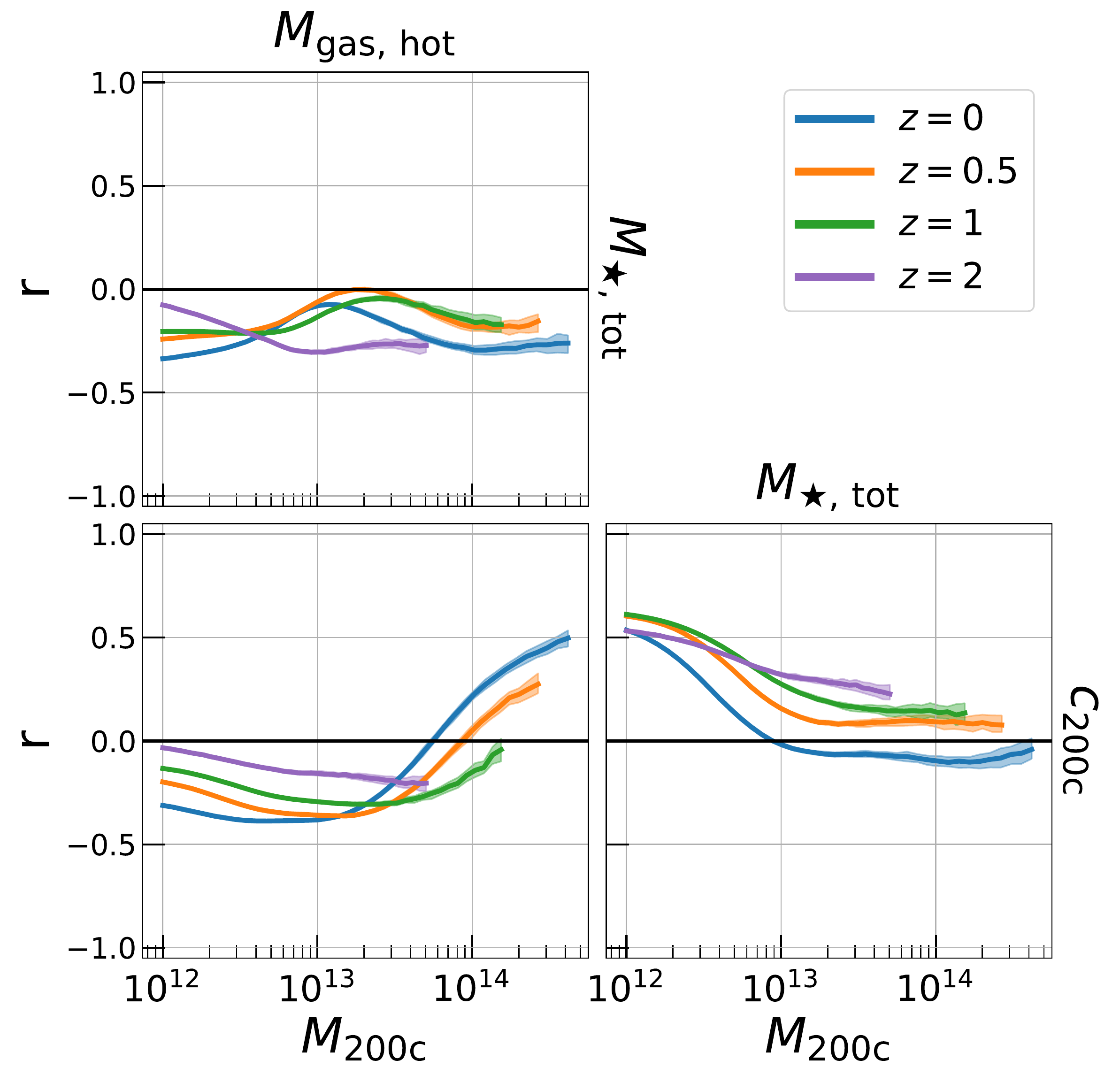}
    \caption{The KLLR local correlations between $\MGasHot$, $\MStar$, and $\ctwohc$ for TNG300 halos. Once again, there are clear mass- and redshift-dependent trends in all the parameters. Note the sign change in the  $\MGasHot$--$\ctwohc$ covariance which reflects how competing effects of star formation efficiency and feedback change with halo formation history. Made using the \texttt{Plot\_Cov\_Corr\_Matrix()} function.}
    \label{fig:TNG_Corr_Matrix}
\end{figure}

\begin{figure}
    \centering
    \includegraphics[width = 0.49\textwidth]{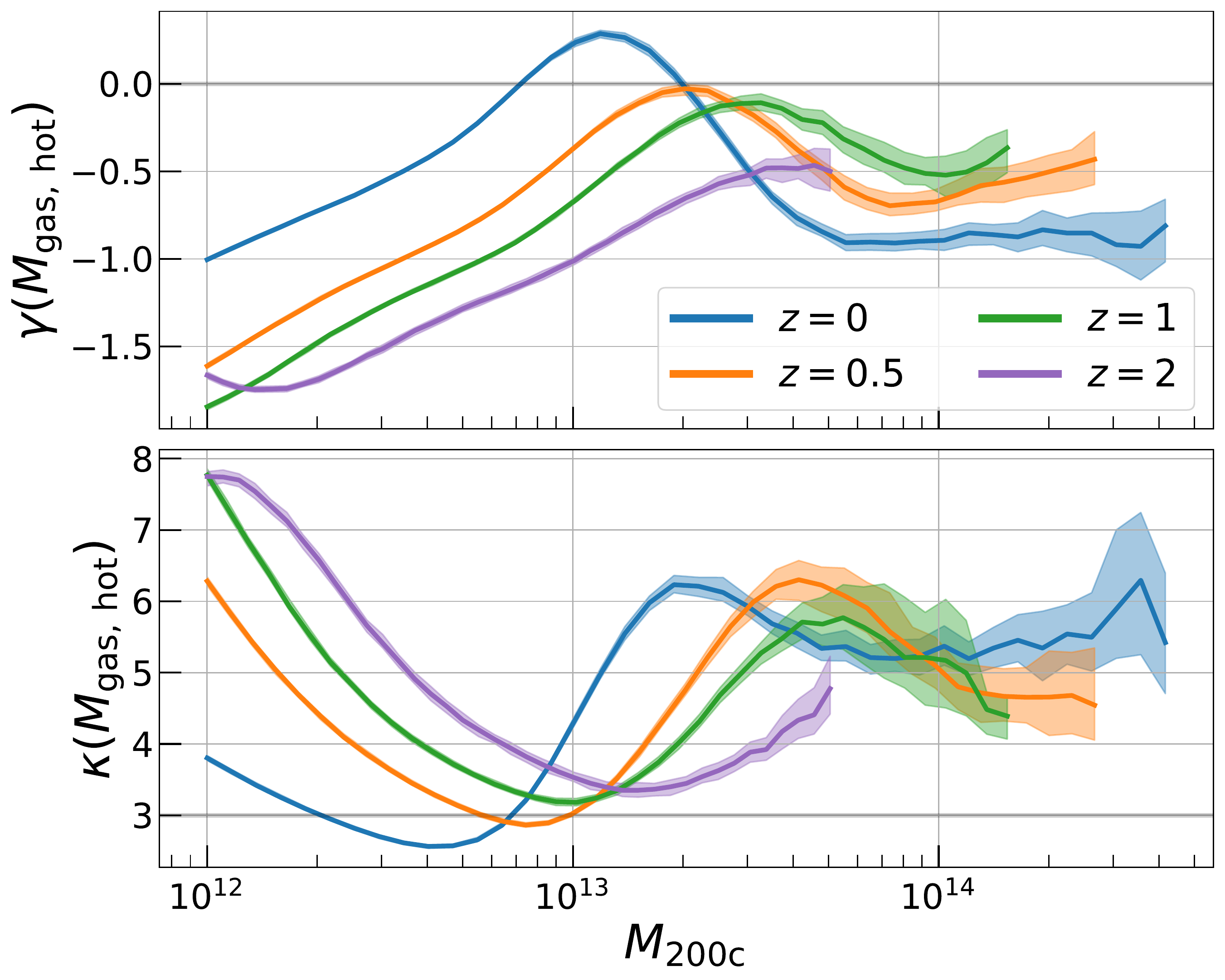}
    \includegraphics[width = 0.49\textwidth]{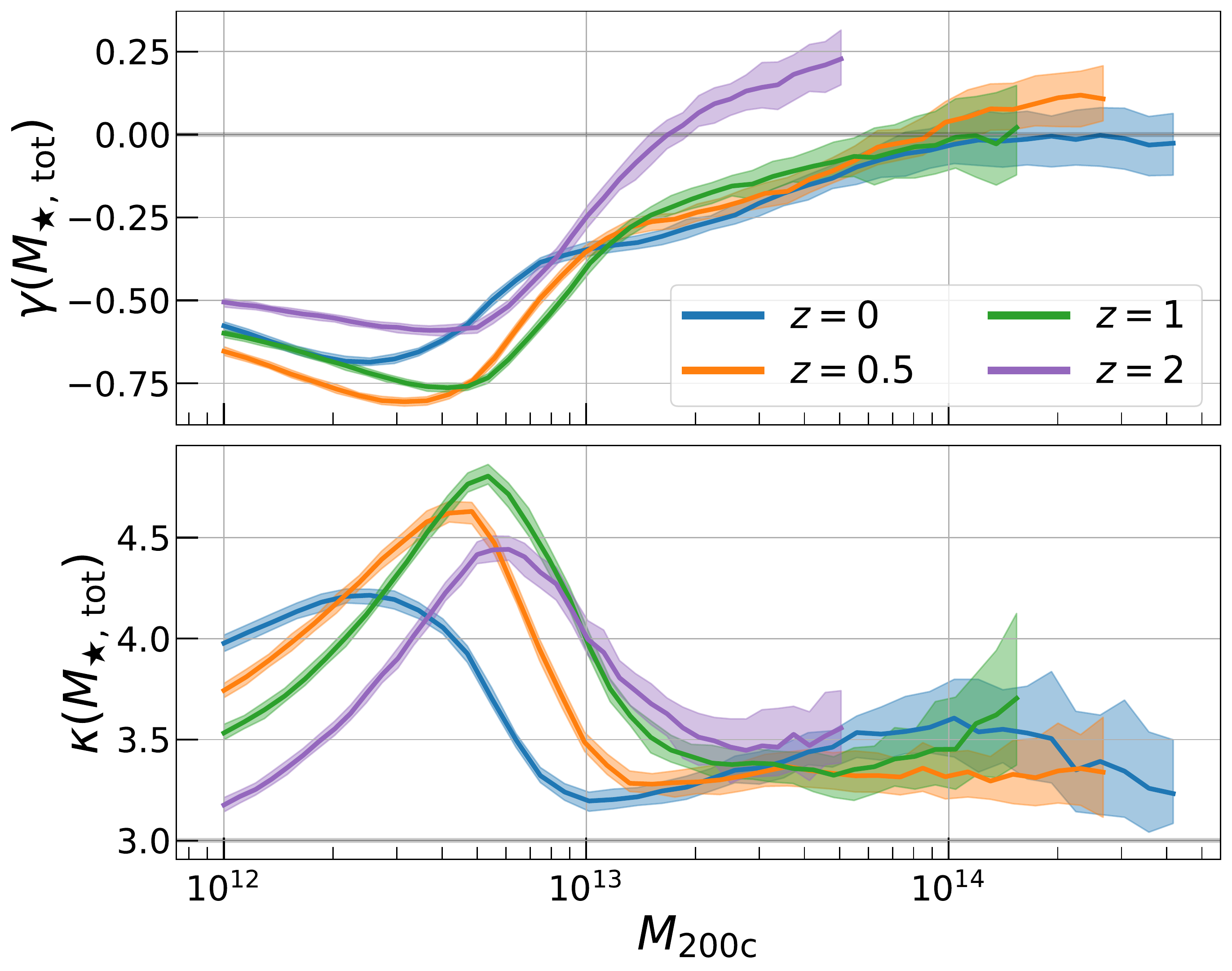}
    \caption{The skewness (top) and kurtosis (bottom) for both $\MGasHot$ (left) and $\MStar$ (right) in the TNG300 halo sample. Horizontal gray lines at $\gamma = 0$ and $\kappa = 3$ are plotted to show the expectation for a Gaussian distribution. Made using the \texttt{Plot\_Higher\_Moments()} function.}
    \label{fig:TNG_diagnosis}
\end{figure}

\begin{figure}
    \centering
    \includegraphics[width = 0.49\textwidth]{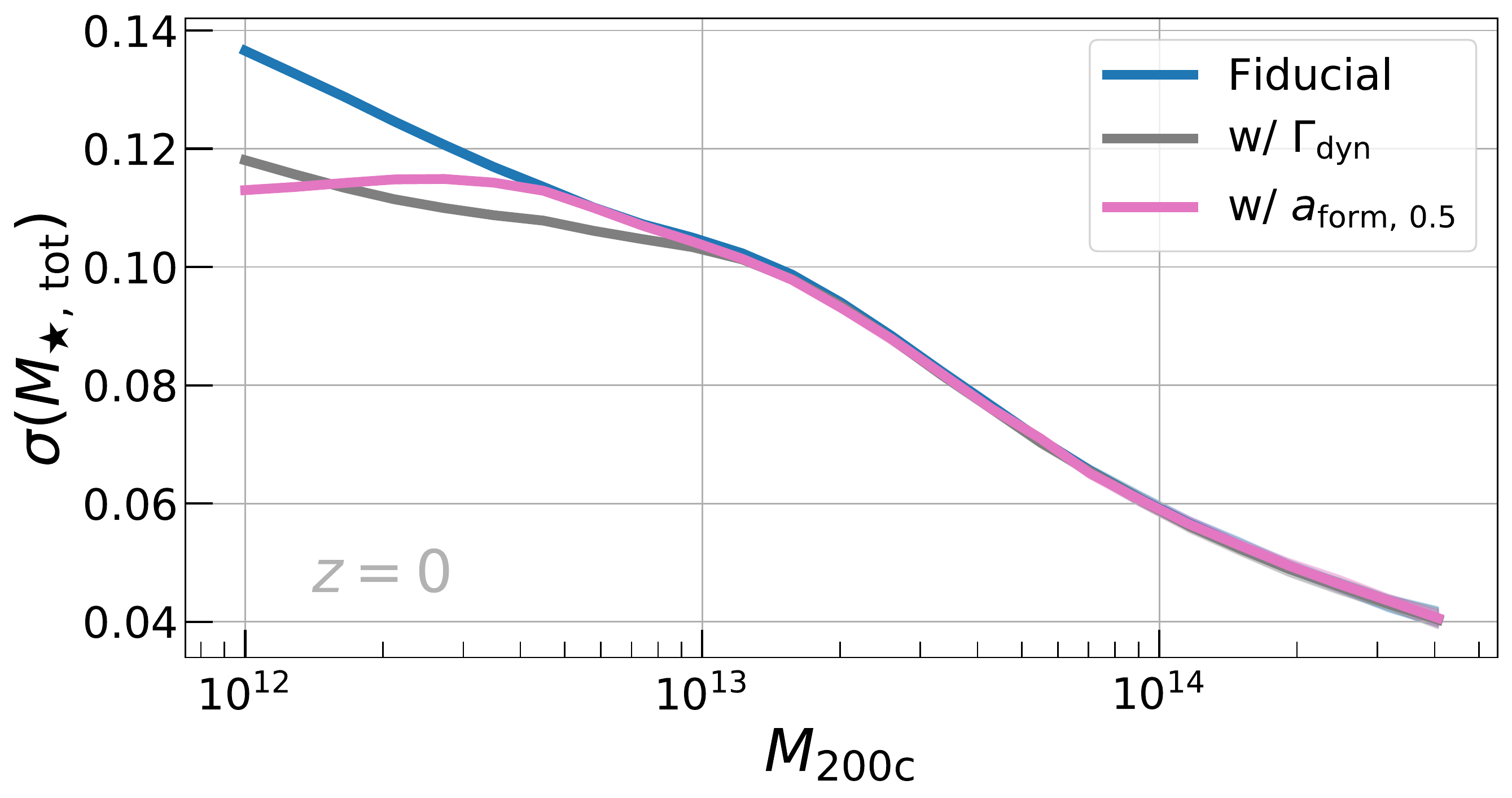}
    \caption{The scatter in $\MStar$ for $z = 0$ TNG300 halos estimated from a univariate regression against $\Mtwohc$ (blue line), along with bivariate regressions that add either the halo accretion rate $\Gamma_{\rm dyn}$ or halo formation time $a_{\rm form,\,0.5}$ to total halo mass. The additional independent variables provide some power in reducing the scatter, but in a significantly mass-dependent way. The 68\% confidence intervals are shown here, but they are smaller than the width of the lines. 
    }
    \label{fig:TNG_Multivariate}
\end{figure}

One of the major challenges in cluster cosmology in the era of large astronomical surveys lies in accuracy of modeling the conditional observable of dark matter halos as a function of their mass and redshift \citep[][]{Allen:2011}.
To optimize the scientific returns of these surveys, it is important to model the Baryonic content of dark matter halos 
as accurately as possible \citep[{\sl e.g.},][]{Mantz:2016-relaxedIII, Costanzi:2019}.
The scaling of cluster observables with halo mass is typically modeled with a linear model and normal scatter in a log-mass -- log-observable space \citep[{\sl e.g.},][]{Vikhlinin:2009,Mantz:2016WtG-V,Mulroy:2019}. But recent results from hyrodynamical simulations has suggested that the simple linear model might be not as accurate as was presumed previously \citep{LeBrun:2017,Farahi:2018,Anbajagane:2020,Anbajagane2021sigma_dm, Anbajagane2021vel_bias}.  
Here we employ simulations to demonstrate that indeed a simple linear model is not adequate as the scaling parameters are mass dependent. We use the halo population of the \textsc{IllustrisTNG} hydrodynamical simulations and show that the scaling parameters -- slope, normalization, covariance -- are all halo mass dependent. Here, we use the log of halo mass\footnote{$\Mtwohc = M(<\Rtwohc$) is defined to satisfy $\Mtwohc = \frac{4\pi}{3}\Rtwohc^3 200 \rho_c(z)$, where $\rho_c(z)$ is the critical density of the Universe at a given epoch. In words, $\Rtwohc$ is the radius within which the mean density is 200 times the critical density, and $\Mtwohc$ is the mass within this radius.}, $\Mtwohc$ , as the scaling variable, $\mu$.

\noindent {\bf Sample:} We employ halos of mass range $10^{12}\,[\msun]$ to $10^{15}\,[\msun]$ derived from the highest resolution TNG300 solution of the IllustrisTNG project\footnote{\url{http://www.tng-project.org/data/}} \citep{Nelson:2018, Springel:2018, Pillepich:2018scaling, Naiman:2018, Marinacci:2018, Nelson:2019}. 
We compute $\MGasHot$ as the total hot gas mass within $R_{\rm 200c}$ of the host halo, where hot gas is determined by a $T > 10^5 K$ temperature cut, and $\MStar$ as the total stellar mass within $R_{\rm 200c}$. We also estimate the halo concentration by fitting an NFW profile \citep{Navarro:1997} to the dark matter density profile, deriving the scale radius, $r_s$, and calculating concentration as $c_{\rm 200c} = R_{\rm 200c}/r_s$.

\noindent  {\bf Results:} Log-halo mass is employed as the scaling variable. For all of our analyses, we use a gaussian kernel in $\log_{10} \Mtwohc$ with width \texttt{kernel\_width = 0.3} dex. Some results after varying this width are shown in Appendix \ref{appx:vary_width}. Figure~\ref{fig:TNG_Fits}, created using the \texttt{Plot\_Fit\_Summary()} function, shows the fitted KLLR scaling parameters --- normalization (top), slope (middle) and scatter (bottom) -- as a function of the scaling variable for dark matter halos in redshifts $0$, $0.5$, $1$, and $2$. As is clearly evident from these results, neither the hot gas mass nor stellar mass follows a global, log-linear scaling with halo mass. These deviations from complete linearity for these two properties are also consistent with previous findings by \citet{Farahi:2018} using the BAHAMAS and MACSIS simulations \citep{Barnes:2017,McCarthy:2017}. 

If massive halos conserve their baryon content, then the total gas and stellar content of halos should be anti-correlated. For the most massive halos that host rich galaxy clusters, this anti-correlation has been detected and quantified both in hydrodynamical simulations \citep{Wu:2015, Farahi:2018} and in empirical data \citep{Farahi:2019}. The TNG halos agree with these results by exhibiting a negative correlation between gas mass and stellar (Figure~\ref{fig:TNG_Corr_Matrix}, made using the \texttt{Plot\_Cov\_Corr\_Matrix()} function), but the absolute value of this anti-correlation varies across simulations \citep{Wu:2015, Farahi:2018} which might suggest that the sub-grid physics plays a role in regulating the covariance between different observables.  

Additionally, the bottom panels of Figure~\ref{fig:TNG_Corr_Matrix} show the correlation of concentration with total gas and stellar mass. 
The correlation between baryonic content and concentration, as opposed to that between hot gas and stellar mass, scales with halo mass and even changes sign.  
This type of scale-dependent signal, readily measured with the KLLR model, suggests that there is complex astrophysics that shapes the relationships among halo properties as a function of halo mass and redshift.

In Figure~\ref{fig:TNG_diagnosis}, made using the \texttt{Plot\_Higher\_Moments()} function, we diagnose the fitted model using a measure of skewness (top) and kurtosis (bottom). There are deviations from the Gaussian assumption (given by $\gamma = 0$ and $\kappa = 3$) for both. The relevance and impact of such deviations vary across science application. For example, if we set our science requirement to be a $<1\%$ bias in the estimated halo mass function due to non-Gaussian scatter \citep{Shaw2010}, then the skewness and kurtosis we see here are close enough to the Gaussian assumption for halos of mass $10^{13}\,\msun$ and above \citep[also see Equation (156) of][]{Weinberg:2013}.
We also notice that scales at which there is a rapid change in the fit parameters coincide with a mass-localized deviation from the normal assumption as captured by the skewness and kurtosis. This might indicate that there are rapid small scale changes in the relations that require smaller kernel widths at around these scales. We show the impact of such changes to the kernel width in Appendix \ref{appx:vary_width}.

The software also performs regression against multiple independent regression variables in addition to the scale variable.  The regression remains conditioned only in the scale dimension, but the variance in the regressed variable will be reduced when there is non-zero covariance with the additional regressed properties.  A potential application of this mode is when multiple observed properties of a population are available from multi-wavelength data samples.  For the case of clusters of galaxies, this may include galaxy stellar mass and weak lensing halo mass estimates from optical/IR photometry, galaxy velocities from spectroscopy,  hot gas mass and temperature from X-ray satellite observations, and hot electron pressure from the Sunyaev-Zel'dovich effect distortion of the cosmic microwave background spectrum. Any one of such measurements could serve as the scale variable, $\boldX$. On the cosmological simulation side, this functionality has also been used to measure the non-linear, multi-property scaling relations of fundamental dark matter halo properties \citep{Anbajagane2021sigma_dm}.

As a theoretical consideration, we choose two additional properties: i) the mass accretion rate, $\Gamma_{\rm dyn}$, which quantifies the rate of at which halo mass grow, $\Mtwohc$, with time \citep{Diemer2017MAR}, and; ii) the formation epoch, $a_{\rm form,\,0.5}$,  the value of the cosmic expansion parameter at which the halo first achieves half its present ($z = 0$, $a=1$) mass \citep{Correa:2015}.  Figure~\ref{fig:TNG_Multivariate} shows the scatter in total stellar mass when regressed on either property in addition to halo mass.  Using either of the properties in addition to halo mass leads to a suppression of scatter at the low-mass end and no effect beyond $\Mtwohc > 10^{13}\, \msun$. 
This example illustrates the necessity of a localized regression model in explaining variance in the response variable. As illustrated in this example the explaining variables and their explaining power can change with the mass-scale. \textsc{KLLR} is designed to capture these scale dependencies in data.

\section{Conclusion}\label{sec:conclusion}

\textsc{KLLR} is an implementation of a scale-dependent, localized linear model that allows the user to uncover scale dependence hidden within traditional linear model approaches. The implementation of the the method is publicly available in a GitHub repository (\href{https://github.com/afarahi/kllr}{\faGithub}).  This python package allows the user to seamlessly perform regression and visualize the resulting parameter behaviors, enabling readily interpretable insights that can inform models of system dynamics. We hope that exposure to this powerful method encourages the astronomy community to go beyond simple linear models, power-law, broken-power law or other parametric models and explore the non-linear trends with data-driven models that will be useful for modeling a variety of astronomical and other data.

\begin{acknowledgments}
The  authors  thank  the  IllustrisTNG  team  for  making their data and catalogs publicly available. DA is supported by the National Science Foundation Graduate Research Fellowship under Grant No. DGE 1746045. AF is supported by the University of Texas at Austin.  AEE acknowledge support from the Leinweber Center for Theoretical Physics. We thank the anonymous referees and editorial staff for historical insights and other feedback that improved the paper's presentation. 
\end{acknowledgments}

\software{\textsc{KLLR} \citep{KLLR},
\textsc{NumPy} \citep{NumPy},
\textsc{Scikit-learn} \citep{Scikitlearn},
\textsc{Pandas} \citep{Pandas}, 
\textsc{Matplotlib} \citep{Matplotlib}.}

\bibliography{mybib}

\appendix

\section{Impact of kernel width} \label{appx:vary_width}

As discussed in the main text, deviations from Gaussianity -- which are implied by skewness $\gamma \neq 0$ and kurtosis $\kappa \neq 3$ -- could indicate real non-Gaussianities in distribution of the data, or they could also arise from using a wider kernel width than that required by the data. In specific, the data may have features that occur on scales smaller than the kernel width, and this in turn causes the fitting procedure to deal with non-Gaussian features.

\begin{figure}
    \centering
    \includegraphics[width = 0.485\textwidth]{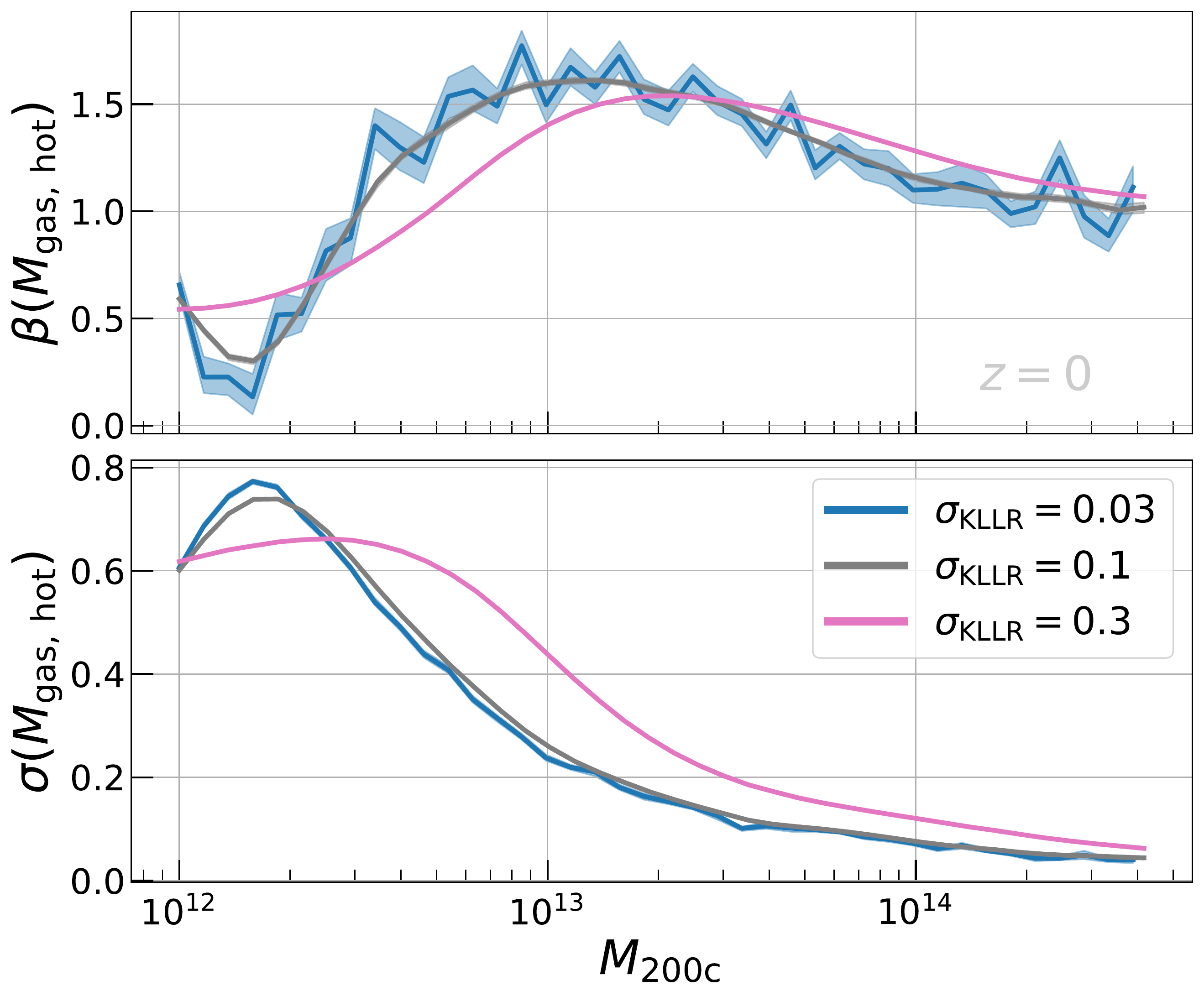}
    \includegraphics[width = 0.502\textwidth]{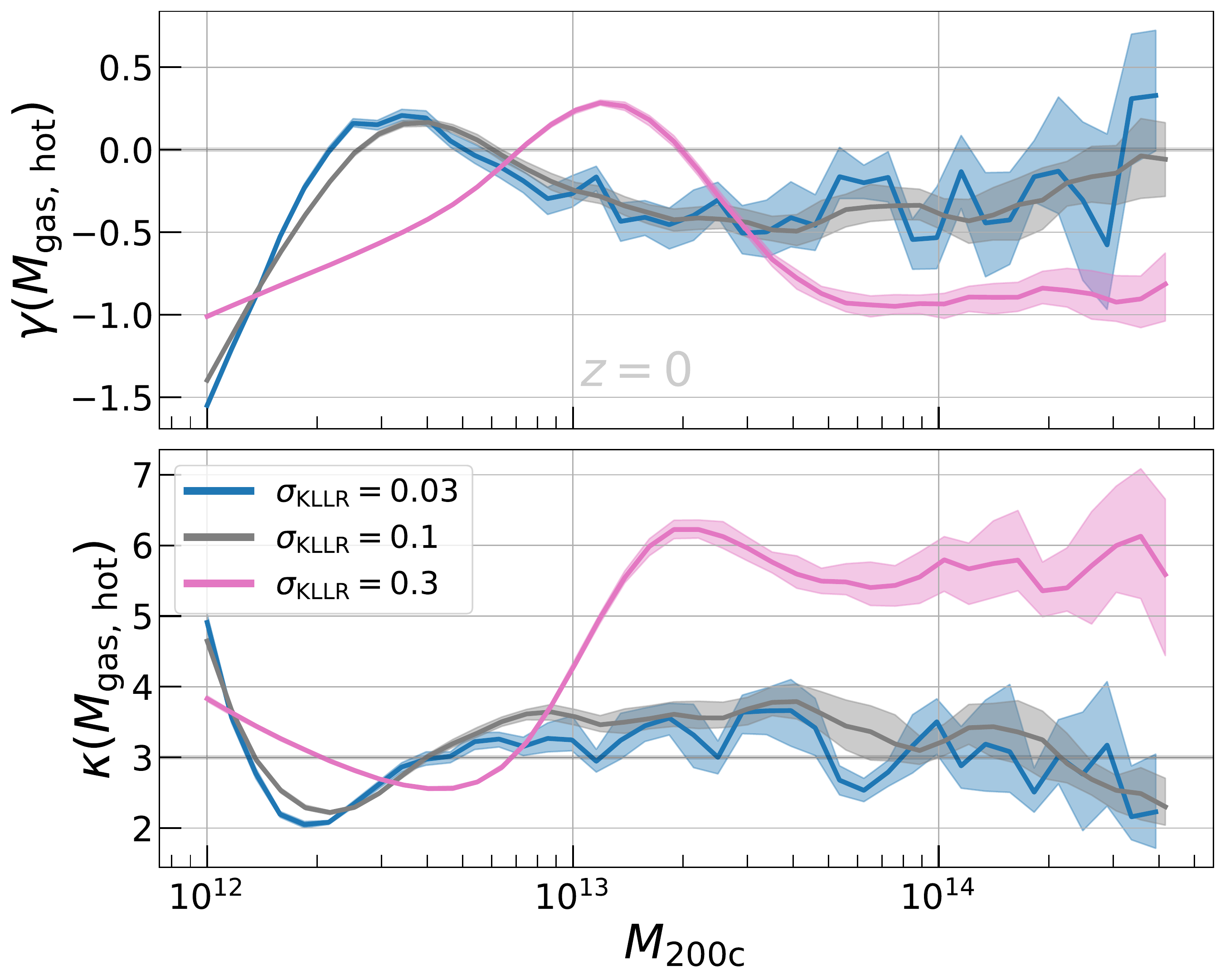}
    \caption{The slope (top left), scatter (bottom left), skewness (top right), kurtosis (bottom right), for the $\MGasHot - \Mtwohc$ relation of TNG300 halos at $z = 0$, estimated using different kernel widths (different colors). Decreasing the kernel width, $\sigma_{\rm KLLR}$ brings the skewness and kurtosis closer to the Gaussian expectations of $\gamma = 0$ and $\kappa = 3$.}
    \label{fig:Vary_width}
\end{figure}

We have shown that the $\MGasHot - \Mtwohc$ relation contains significant non-Gaussianities (figure \ref{fig:TNG_Fits}). In figure \ref{fig:Vary_width}, we vary the kernel width used in our KLLR parameter estimation and show the fit parameters (left panel) and the diagnostic metrics (right panel). Our original choice of $\sigma_{\rm KLLR} = 0.3$ is likely too wide, as reducing the width to $\sigma_{\rm KLLR} = 0.1$ brings the skewness closer to $\gamma = 0$ while the kurtosis is now statistically consistent with $\kappa = 3$ for most of the halo mass scales. However, reducing the width even further to $\sigma_{\rm KLLR} = 0.03$ does not alter the KLLR parameters nor the moments, but simply results in a noisier estimate as evidenced by the jagged features in the lines.

\end{document}